\documentclass[prd,nofootinbib,floats,superscriptaddress,eqsecnum,tightenlines,11pt]{revtex4}

\usepackage{hyperref}
\usepackage{graphicx}
\usepackage{amsmath,amssymb,amsfonts,amsthm,latexsym,stmaryrd}
\usepackage{marginnote}
\usepackage{color}

\def\beq{\begin{equation}}
\def\eeq{\end{equation}}
\newcommand{\bea}{\begin{eqnarray}}
\newcommand{\eea}{\end{eqnarray}}
\def\bi{\begin{itemize}}
\def\ei{\end{itemize}}
\def\ba{\begin{array}}
\def\ea{\end{array}}
\def\bfig{\begin{figure}}
\def\efig{\end{figure}}

\def\tA{\hat A} 
\def\B{{\cal B}}
\def\a{\alpha}
\def\f{f}

\def\aa{\alpha_2}
\def\ab{\alpha_1}
\def\ac{\alpha_3}
\def\ad{\alpha_4}
\def\ae{\alpha_5}
\def\ba{\beta_1}
\def\bb{\beta_2}
\def\tba{\tilde\beta_1}
\def\tbb{\tilde\beta_2}
\def\ka{\kappa_2}
\def\kb{\kappa_1}
\def\kc{\kappa_3}
\def\kd{\kappa_4}
\def\ke{\kappa_5}
\def\tka{\tilde\kappa_1}
\def\tkb{\tilde\kappa_2}
\def\tkc{\tilde\kappa_3}

\def\ga{\gamma_2}
\def\gb{\gamma_1}
\def\gc{\gamma_3}

\def\An{A_*}
\def\dotAn{\dot A_*}
\def\A{{\cal A}}
\def\V{{\cal V}}
\def\M{{\cal M}}
\def\R{{}^{(4)}\!R}
\def\Rs{{}^{(3)}\!R}
\def\D{D}

\def\n{n} %number of degrees of freedom

\newcommand{\Gfour}{G_4{}}
\newcommand{\Ffour}{F_4{}}

\begin{document}

\title{Degenerate higher derivative theories beyond Horndeski: evading  the Ostrogradski instability}
\author{David  Langlois}
\email{langlois@apc.univ-paris7.fr}
\affiliation{APC, Astroparticule et Cosmologie, CNRS--Universit\'e Paris Diderot, 10 rue Alice Domon et L\'eonie Duquet,  75205 Paris Cedex 13, France}
\author{Karim Noui}
\email{karim.noui@lmpt.univ-tours.fr}
\affiliation{Laboratoire de Math\'ematiques et Physique Th\'eorique, Universit\'e Fran\c cois Rabelais, Parc de Grandmont, 37200 Tours, France}
\affiliation{APC, Astroparticule et Cosmologie, CNRS--Universit\'e Paris Diderot, 10 rue Alice Domon et L\'eonie Duquet,  75205 Paris Cedex 13, France}

\date{\today}

\begin{abstract}
Theories with higher order time derivatives generically suffer from ghost-like instabilities, known as Ostrogradski instabilities. This fate can be avoided by considering ``degenerate'' Lagrangians,  whose kinetic matrix  cannot be inverted, thus leading to constraints between canonical variables and 
a reduced number of physical degrees of freedom. In this work, we derive in a systematic way the degeneracy conditions  for scalar-tensor theories that depend quadratically  on second order derivatives of a scalar field. We thus obtain a classification of all degenerate theories within this class of scalar-tensor theories.
    The quartic Horndeski Lagrangian and its extension beyond Horndeski belong to these degenerate cases.
 We also identify new families of  scalar-tensor theories with the  property that they are degenerate despite the nondegeneracy of the purely scalar part of their Lagrangian.  
\end{abstract}

\maketitle

\section{Introduction}
Theories of modified gravity have attracted considerable attention in the last few years (see e.g. \cite{Joyce:2014kja,Berti:2015itd,Koyama:2015vza} for recent reviews). The main motivation that drives their study is to  find an explanation for  the present cosmic acceleration, even if  exploring alternative theories of gravitation is also very instructive from a more fundamental point of view.  
Many theories of modified gravity are  constructed by simply introducing a scalar field in addition to the  usual tensor modes of general relativity, hence their generic name of scalar-tensor theories. The latest studies of dark energy have tried to encompass very large classes of scalar-tensor theories (see e.g. \cite{Gleyzes:2013ooa,Gleyzes:2014rba,Gleyzes:2015pma}   for a recent approach that unifies the treatment of  single scalar field models of dark energy). 

One way to enlarge the traditional scalar-tensor theories is to allow for the presence of higher order derivatives in the Lagrangian. However, this possibility is severely restricted  in order to avoid disastrous instabilities. Indeed, 
 according to Ostrogradski's theorem, nondegenerate Lagrangians with higher order time derivatives\footnote{For a single variable $q$, a Lagrangian of the form $L(q, \dot q, \ddot q)$ is nondegenerate if $\partial^2 L/\partial\ddot q^2\neq 0$. Multi-variable Lagrangians will be discussed in the main text.}  lead to ghost-like instabilities, also known as Ostrogradski instabilities~\cite{Ostrogradsky}.
Such Lagrangians yield higher order equations of motion, which require more initial conditions than in usual dynamical systems. This translates, in the Hamiltonian formulation, into the appearance of  an extra degree of freedom, with a Hamiltonian that depends linearly on one canonical momentum and is thus (kinetically) unbounded from below. 
  
This lethal fate can however be avoided for special Lagrangians. 
Well-known examples are  the galileon models~\cite{NRT}, which lead to second order equations of motion despite the presence of higher order derivatives in their Lagrangian.  
For 4D scalar-tensor theories, i.e. including both a scalar field $\phi$ and a four-dimensional dynamical metric $g_{\mu\nu}$,  the 
most general Lagrangians
leading to second order equations of motion for $\phi$ and $g_{\mu\nu}$ were obtained by Horndeski~\cite{horndeski}. These models were 
later rediscovered  in   constructing  general covariant extensions of the galileons  that lead to  at most second order equations of motion, with the preconception that this requirement was necessary to avoid Ostrogradski ghosts~\cite{Deffayet:2011gz,Kobayashi:2011nu}. 
Hordenski's  theories  can be described by   linear combinations of the following Lagrangians,
\begin{align}
L_2^{\rm H} & \equiv  G_2(\phi,X)\;, \qquad \qquad 
L_3^{\rm H}  \equiv  G_3(\phi, X) \, \Box \phi \;, \label{L3} \\
L_4^{\rm H} & \equiv G_4(\phi,X) \, {}^{(4)}\!R - 2 G_4{}_{,X}(\phi,X) (\Box \phi^2 - \phi^{ \mu \nu} \phi_{ \mu \nu})\;, \\
L_5^{\rm H} & \equiv G_5(\phi,X) \, {}^{(4)}\!G_{\mu \nu} \phi^{\mu \nu}  +  \frac13  G_5{}_{,X} (\phi,X) (\Box \phi^3 - 3 \, \Box \phi \, \phi_{\mu \nu}\phi^{\mu \nu} + 2 \, \phi_{\mu \nu}  \phi^{\mu \sigma} \phi^{\nu}_{\  \sigma})  \,,
\end{align}
where we have used the notation $\phi_\mu\equiv \nabla_\mu \phi, \,  \phi_{\mu\nu}\equiv \nabla_{\nu}\nabla_\mu \phi$ and $X\equiv \nabla_\mu\phi\nabla^\mu\phi$, and a comma denotes a partial derivative with respect to the argument.

Recently it was realized  that requiring second order equations of motion is in fact not mandatory and  extensions of Horndeski's quartic and quintic Lagrangians  were proposed in \cite{Gleyzes:2014dya,Gleyzes:2014qga} (an earlier example of theory beyond Horndeski was constructed in \cite{Zumalacarregui:2013pma} via the use of disformal transformations). 
These additional Lagrangians can be written in the form
\begin{align}
L_4^{\rm bH}& \equiv F_4(\phi,X) \epsilon^{\mu\nu\rho}_{\ \ \ \ \sigma}\, \epsilon^{\mu'\nu'\rho'\sigma}\phi_{\mu}\phi_{\mu'}\phi_{\nu\nu'}\phi_{\rho\rho'}\;, \label{L4} \\
L_5^{\rm bH}&\equiv F_5 (\phi,X) \epsilon^{\mu\nu\rho\sigma}\epsilon^{\mu'\nu'\rho'\sigma'}\phi_{\mu}\phi_{\mu'}\phi_{\nu \nu'}\phi_{\rho\rho'}\phi_{\sigma\sigma'} \label{L5bH}\,,
\end{align}
where   $\epsilon_{\mu \nu \rho \sigma }$ is the totally antisymmetric Levi-Civita tensor. 
They  lead  to equations of motion that are third order in time derivatives. 
Interestingly, these extensions beyond Horndeski can also be recast as generalizations of the ``John'' and ``Paul'' terms  of the Fab Four~\cite{Charmousis:2011bf}, where the corresponding two arbitrary functions of $\phi$ acquire a dependence on  $X$ as well~\cite{Babichev:2015qma}.
These theories beyond Horndeski lead to a whole range of new phenomena, which have been recently investigated in several works (see e.g. \cite{Fasiello:2014aqa,Kobayashi:2014ida,Koyama:2015oma,Saito:2015fza,DeFelice:2015isa,Tsujikawa:2015mga,Tsujikawa:2015upa,Sakstein:2015zoa}).

Evading the Ostrogradski ghost usually requires to work with a  ``degenerate'' Lagrangian\footnote{Another approach, familiar in the context of effective field theory, consists in simply discarding  Ostrogradski instabilities when they arise from the {\it perturbative} part of the Lagrangian (see e.g. \cite{Burgess:2014lwa}).}.  In this sense,  flat spacetime galileons can be seen as degenerate theories. Degeneracy can also involve several variables simultaneously and is the main focus of the present work. 
To be more specific, we  define a degenerate theory as follows. After introducing 
auxiliary variables to replace the second order time derivatives of the Lagrangian by first order time derivatives, the Lagrangian is said to be  degenerate if the kinetic matrix (composed of the coefficients of the kinetic terms) is degenerate. As we show in this paper, Horndeski theories are degenerate in a trivial way: there is no mixing in the kinetic matrix between the higher order sector (scalar field) and the metric sector. 
By contrast, the extension beyond Horndeski is
characterized by a nontrivial degeneracy, which involves a mixing between the two sectors and explains why the equations of motion are higher order even if the system remains degenerate.

The central purpose of this paper is  to derive, in a systematic way, the degeneracy conditions for  higher derivative scalar theories coupled to gravity. For pedagogical reasons, we first introduce a toy model 
with properties that are very similar to those encountered in the scalar-tensor theories we investigate later. 
This toy  model is sufficiently simple that  the equations of motion and the Hamiltonian formulation can be easily derived and fully analysed.  

We then turn to a large class of scalar tensor theories with  Lagrangians that depend quadratically on the second derivatives of the field, i.e.  $\phi_{\mu\nu}$, while the dynamics of the gravitational sector  is described by the Ricci scalar multiplied by an arbitrary function of $\phi$ and $X$. This class of theories depends on six arbitrary functions of $\phi$ and $X$, and includes the quartic Horndeski and beyond Horndeski terms, i.e. $L_4^{\rm H}$ and $L_4^{\rm bH}$. 
Familiar terms like a standard kinetic term or a potential, or more generally any combination of $L_2^{\rm H}$ and $L_3^{\rm H}$, can of course be added, but we do not consider them explicitly as they do not modify the degeneracy properties of the Lagrangian.

For the class of theories described above, we are able to derive the degeneracy conditions that must be satisfied by the arbitrary functions of the Lagrangian. This enables us to classify all degenerate Lagrangians in this class. In particular, we find that a combination of $L_4^{\rm H}$ and $L_4^{\rm bH}$ is degenerate, as expected. We also identify three other families of degenerate Lagrangians, whose purely scalar part  is nondegenerate.  

We also consider separately the quintic Lagrangian beyond Horndeski, $L_5^{\rm bH}$, whose analysis is more involved since its dependence on $\phi_{\mu\nu}$  is now  cubic. We show that $L_5^{\rm bH}$  is degenerate and the corresponding null eigenvector of its kinetic matrix is the same as that of $L_4^{\rm bH}$, which implies that $L_4^{\rm bH}+L_5^{\rm bH}$ is also degenerate. However, we  find that the combination of $L_4^{\rm bH}$, $L_4^{\rm bH}$ and $L_5^{\rm bH}$ leads to a {\it nondegenerate} Lagrangian in general, which means that an Ostrogradski instability is expected in this case.

The plan of our paper is the following. The next section is devoted to the full analysis of our toy model.  We then present the class of models we investigate, which include the quartic Lagrangians $L_4^{\rm H}$ and  $L_4^{\rm bH}$  as particular cases. The  section that follows explains how to derive  the degeneracy conditions. This enables us to fully classify the degenerate theories in the subsequent section.  We then consider  the particular case of $L_5^{\rm bH}$. We finally summarize our results and conclude. 
In addition, we present  a discussion on the unitary gauge and the degeneracy conditions  in an Appendix.

\section{A toy model}
In order to illustrate how the Ostrogradski instability can be circumvented within {\it degenerate} theories, despite the presence of higher order time derivatives in the equations of motion, let us consider and study in detail a  simplified toy model. The degeneracy properties that we discuss in this section will be quite similar to those of the scalar-tensor theories  considered in the rest of the paper. 

\subsection{Higher derivative Lagrangian}
Our toy  model describes a point particle system with higher derivatives, coupled to $\n$ regular degrees of freedom.   Denoting the respective variables by $\phi(t)$ and $q^i(t)$ 
 ($i=1,\dots, \n$), their coupled dynamics is governed by a Lagrangian of the form
\beq\label{toy model action}
L=\frac12a\, \ddot \phi^2+  \frac 12 k_0 \dot \phi^2 + \frac12 k_{ij} \dot q^i \dot q^j + b_i \, \ddot \phi \, \dot q^i+c_i \, \dot \phi \,  \dot q^i
- V(\phi,q) \,.
\eeq
In this simplified framework, we assume that $a$, $b_i$, $c_i$, $k_0$ and $k_{ij}$ are constant but the model could easily be extended to arbitrary functions of $\phi$. 
The  coupling between the variable $\phi$ and the regular degrees of freedom $q^i$ is governed by 
  two interaction terms. In particular, the interaction term proportional to $b_i$  generates third order derivatives
in the equations of motion for $\phi$ and $q^i$, which read respectively
\begin{eqnarray}
a \ddddot \phi- k_0 \ddot \phi  + b_i \dddot q^i  - c_i \ddot q^i  - V_\phi & = &0\; ,\label{eq for Phi}\\
k_{ij} \ddot q^j+b_i \dddot \phi + c_i \ddot \phi+  V_i   & = & 0 \label{eq for q} \,,
\end{eqnarray}
where $V_i\equiv\partial V/\partial q^i$ and $V_\phi\equiv\partial V/\partial \phi$.
In general,  these equations involve an  extra degree of  freedom, corresponding to an Ostrogradski ghost.   However, as we show in more details below, there exist cases where the ghost can be avoided even if the equations of motion feature higher order time derivatives. 

\subsection{An equivalent formulation}
To compute the number of degrees of freedom (either in  the Lagrangian or Hamiltonian frameworks), it is convenient to reformulate the theory in a way that eliminates explicit higher order time derivative in the Lagrangian. For that purpose, we simply replace $\dot \phi$ by a new variable $Q$ in (\ref{toy model action}) and we add a ``constraint"  which imposes
indeed that $Q$ is the time derivative of $\phi$. Thus, we introduce the new Lagrangian
\beq\label{reformulated toy model}
L=\frac12 a\, \dot Q^2 + \frac12 k_{ij} \dot q^i \dot q^j + \frac12 k_0 Q^2 - V(\phi,q) + ( b_i \dot Q + c_i Q)\dot q^i - \lambda (Q-\dot \phi)\,,
\eeq
where $Q$ and $\lambda$ are two new variables. 

To verify that  this action is equivalent
to the original one (\ref{toy model action}), we derive  the equations of motion, which read
\begin{eqnarray}
 a\ddot Q + b_i \ddot q^i & = & c_i \dot q^i +  k_0 Q - \lambda \, , \label{eq for Q} \\
b_i \ddot Q + k_{ij} \ddot q^j & = &  -  V_i -  c_i \dot Q\,, \label{eq modified for q}\,\\
 \dot \phi \; = \; Q \;\;&\; \text{and} \;&\;\; \dot \lambda  =- V_\phi   \label{extra eq}\;,
\end{eqnarray}
and check explicitly that  there are indeed equivalent to the system (\ref{eq for Phi})-(\ref{eq for q}). 

We now introduce the kinetic matrix, i.e. the symmetric matrix  that contains the coefficients of the terms quadratic in time derivatives in the new Lagrangian (\ref{reformulated toy model}):
\begin{eqnarray}
M = \left(
 \begin{array}{cc}
 a & {b_j} \\
{b_i} & {k_{ij}}
 \end{array}
 \right) \,.
 \label{kinetic_matrix}
 \end{eqnarray}
As we will see, this matrix plays a crucial r\^ole in the determination of the number of degrees of freedom.
 
 If $M$ is invertible, the equations (\ref{eq for Q}-\ref{eq modified for q}) enable us to express the second order  derivatives $\ddot Q$ and $\ddot q^i$ in terms of  up to first order derivative quantities. Together with the equations in (\ref{extra eq}), the differential system thus requires  initial conditions for $Q$, $\dot Q$, $q^i$, $\dot q^i$, $\lambda$ and $\phi$, i.e. $2(\n+2)$ initial conditions. This means that  the system describes ($\n+2$) degrees of freedom, which includes the extra degree of freedom associated with the Ostrogradski ghost.
 In conclusion,  when  the kinetic matrix $M$ is invertible, the system (\ref{reformulated toy model}) admits  a ghost and provides a typical illustration of the Ostrogradski instability. The same conclusion can also be reached from the Hamiltonian point of view, with a precise analysis of the constraints and the counting of the number of degrees of freedom. This will be the purpose of the last subsection. 

\subsection{Degeneracy: eliminating the extra degree of freedom} 
The presence of an extra degree of freedom can be avoided by  imposing that the kinetic matrix $M$ is degenerate.  We also require that this degeneracy arises from the $\phi$ sector and its coupling to the $q^i$, not from the $q^i$ sector alone, which means that we assume the matrix $k_{ij}$ to be invertible.
By writing the determinant of $M$ in the form 
\begin{eqnarray}
\text{det}(M)=\text{det}(k) \left(a-b_i b_j (k^{-1})^{ij} \right) \,, \label{detM}
\end{eqnarray} 
one finds that the degeneracy of the kinetic matrix $M$ is expressed by  the algebraic relation
\begin{eqnarray}\label{degeneracy condition}
a- b_i \, b_j \, (k^{-1})^{ij} \; = \; 0 \,. \label{D}
\end{eqnarray}
An obvious way to make $M$ degenerate is to choose $a=0$ and $b_i=0$. In this trivial case, all the higher  order derivatives disappear in the original Lagrangian and the system describes $\n+1$ degrees of freedom as usual. This choice also implies that the equations of motion are second order. 

Let us now turn to more interesting situations where the degeneracy is nontrivial, i.e. when the degeneracy condition (\ref{degeneracy condition}) is satisfied with $b_i\neq 0$.
Although the associated equations of motion  involve higher order time derivatives (up to fourth order if $a\neq 0$, third order otherwise),  the degeneracy guarantees that there is no extra degree of freedom. 
 
In order to see this, let us introduce the vector 
\begin{eqnarray}
v = \left( \begin{array}{c} v^0\\ v^i  \end{array}\right) = \left( \begin{array}{c} -1\\ (k^{-1})^{ij}b_j  \end{array}\right) \,,\label{defv}
\end{eqnarray}
which is a generator of the one-dimensional  kernel of the matrix $M$. Projecting the system (\ref{eq for Q})-(\ref{eq modified for q})  in the direction $v$ eliminates all second order derivatives and gives
\begin{eqnarray}\label{new}
c_i (\dot q^i + v^i \dot Q) + k_0 Q + v^i V_i = \lambda \,.
\end{eqnarray}
The above equation suggests  to work with the variables
$x^i\equiv q^i + v^i Q$,  instead of the $q^i$.  In terms of these new variables, the equations of motion of the dynamical system  simplify into
\begin{eqnarray}
&&c_i \dot x^i + k_0 Q + v^i V_i = \lambda\,,  \label{new2}\\
&&k_{ij}\ddot x^j + c_i \dot Q + V_i = 0 \label{new3} \,.
\end{eqnarray}
Taking the time derivative of the first equation (\ref{new2}) and using (\ref{extra eq}) to eliminate $\dot\lambda$ and $Q$, we finally obtain the equivalent dynamical system 
\begin{eqnarray}
 (k_0 - v^i v^j  V_{ij}) \ddot \phi  + c_i \ddot x^i & = & - (v^i  V_{ij})\dot x^j  - (v^i V_{i\phi}) \dot \phi - V_\phi  \label{eq for phi modified}\\
  c_i \ddot \phi  + k_{ij}\ddot x^j & =  & - V_i   \,,\label{eq final for q}
\end{eqnarray}
where $V_{ij}\equiv \partial V_i/\partial q^j=V_{ji}$ and $V_{i\phi}=\partial V_i/\partial\phi=V_{\phi i}$ (the potential $V$ and its derivatives depend on $\phi$, $\dot\phi$ and $x^i$ via the substitution 
$q^i=x^i-v^i \dot \phi$).

We have thus  obtained a second order system for the variables $x^i$ and $\phi$,  which means that the theory  generically requires $2(\n+1)$ initial conditions to be solved. Note that the new system is itself degenerate when the new kinematic matrix
 \begin{eqnarray}
\tilde{M} = \left(
 \begin{array}{cc}
 k_0 -v^iv^j V_{ij} & c_i \\
 c_i & k_{ij}
 \end{array}
 \right)
 \end{eqnarray}
 is not invertible. This occurs  if its determinant,   
 \begin{eqnarray}
 \text{det}(\tilde{M}) = \Delta \, \det{k} \,\;\;\; \text{with} \;\;\;\; \Delta=k_0 -v^iv^j V_{ij} -(k^{-1})^{ij} c_i c_j \,,\label{Delta}
 \end{eqnarray}
 vanishes, i.e.  if $\Delta=0$ since $k$ is invertible.
 A careful analysis of this particular case would show that, in this case, the theory admits in fact fewer physical degrees of freedom.
 From now on, we will assume that the potential $V$ is generic and  that $\Delta$ does not vanish.

\subsection{Hamiltonian analysis}
The Hamiltonian formulation is certainly the most rigorous framework to count the number of physical degrees of freedom and to study the stability  of the system. To complete the previous
Lagrangian analysis, we now perform  the canonical analysis of the theory.

In the Hamiltonian framework, the configuration variables and  their respective conjugate momenta satisfy the Poisson brackets:
\begin{eqnarray}
\{P,Q\} = 1 \;\;, \;\; \{p_i,q^j\}=\delta_i^j \;\;,\;\; \{\pi_\phi,\phi\}=1 \,,
\end{eqnarray}
while the other Poisson brackets vanish. For simplicity, we do not consider here $\lambda$ as an independent variable, but we simply identify it with the conjugate momentum $\pi_\phi$, as it follows from the Lagrangian (\ref{reformulated toy model}) that $\pi_\phi\equiv \partial L/\partial \dot\phi=\lambda$.  Similarly, one finds that the momenta $P$ and $p_i$ are related to $\dot Q$ and $\dot q^i$ by 
\begin{eqnarray}
\label{Pi}
\left( \begin{array}{c} P \\ p_i \end{array} \right) = M \left( \begin{array}{c} \dot{Q} \\ \dot{q}^j \end{array} \right) + 
\left( \begin{array}{c} 0 \\ Q \, {c}_i \end{array}  \right)\,,
\end{eqnarray}
where $M$ is the kinetic matrix (\ref{kinetic_matrix}).

\subsubsection{Non-degenerate case: Ostrogradki's ghost}
 When $M$ is invertible, it is possible to invert the system (\ref{Pi}) and to express the velocities $\dot Q$ and $\dot q^i$ in terms of the momenta $P$ and $p_i$.
     The Hamiltonian is thus given by
\begin{eqnarray}
H & = & P\dot Q + p_i \dot{q}^i + \pi_\phi \dot \phi - L \\
& = & \frac{1}{2} \left( \begin{array}{cc} P ,& p_i - Q\, c_i \end{array}\right) M^{-1} \left( \begin{array}{c} P \\ p_j - Q\, c_j\end{array} \right) + V(\phi,q) -\frac{1}{2}k_0 Q^2 + \pi_\phi Q \,.
\end{eqnarray}
The Hamiltonian $H$ is a function of the $2(\n+2)$ canonical variables: $(Q,q^i,\phi)$ and $(P,p_i,\pi_\phi)$, corresponding to $\n+2$ degrees of freedom, as obtained in the Lagrangian analysis. Moreover,  one observes that the Lagrangian is linear in $\pi_\phi$, which makes the Hamiltonian unbounded from below.  This is the characteristic signature of Ostrogradski's instability.

\subsubsection{Degenerate case}
The only possibility to avoid the Ostrogradski ghost is to assume that $M$ is degenerate, i.e.  that the condition (\ref{degeneracy condition}) is satisfied. Note that such a condition
does not necessarily imply that the coefficient $a$ vanishes. 

An immediate consequence of the degeneracy is  the existence of a primary constraint  relating the canonical momenta, which reads
\begin{eqnarray}
\Omega =  v^i(p_i-Qc_i) - P \, \approx \, 0.
\label{Omega}
\end{eqnarray}
As usual, we use the notation $\approx$  to denote  weak equality in phase space. 
We then introduce the canonical  Hamiltonian  defined by
\begin{eqnarray}\label{def Hamilton}
H= P\dot Q+p_i \dot q^i + \pi_\phi \dot \phi- L \,.
\end{eqnarray}
After some straightforward manipulations to eliminate the velocities, and taking into account the primary constraint (\ref{Omega}), one finds that the expression of the total Hamiltonian  in terms of the canonical variables is given by
\begin{eqnarray}
H_T = \frac{1}{2} (k^{-1})^{ij} (p_i - Q c_i)( p_j - Qc_j) - \frac{1}{2} k_0 Q^2 + V(\phi,q) + \pi_\phi Q + \mu \, \Omega \,,
\end{eqnarray}
where $\mu$ is a Lagrange multiplier enforcing the primary constraint (\ref{Omega}).

The invariance under time evolution of the constraint $\Omega$ leads to the secondary constraint
\begin{eqnarray}
\Psi = \dot \Omega=\{\Omega , H_T\} = c_i (k^{-1})^{ij} (p_j-Qc_j) + k_0 Q + v^i V_i - \pi_\phi \; \approx \; 0 \,.
\end{eqnarray}
To see whether time evolution of $\Psi$ leads to tertiary constraint, it is sufficient to compute the Poisson bracket between the 
primary and secondary constraints,
\begin{eqnarray}
\{ \Omega , \Psi\} = k_0 - v^i v^j V_{ij} - (k^{-1})^{ij}c_ic_j =  \Delta \;,
\end{eqnarray}
where one recognizes the expression $\Delta$ that already  appeared in the Lagrangian framework (see Eq.~(\ref{Delta})). As before, we leave aside the special case $\Delta=0$ (which would further reduce the number of physical degrees of freedom). 

In the generic case where $\Delta\neq 0$, the analysis stops here because imposing $\dot\Psi=0$ simply fixes the Lagrange multiplier $\mu$ without generating any new constraint. 
We have thus obtained a Hamiltonian system with a $2(\n+2)$-dimensional phase space restricted by two second-class constraints $\Omega$ and $\Psi$. This implies that the number of physical degrees of freedom is only $\n+1$. There is no extra degree of freedom in this degenerate case. 

Moreover, one can construct  the physical phase space spanned by the variables $(q^i,p_i;Q,\phi)$, with  a Poisson algebra defined from the Dirac bracket (see e.g. \cite{HT})
\begin{eqnarray}
\{F,G\}_D = \{F,G\} - \frac{1}{\Delta} \left( \{F,\Psi\}\{\Omega,G\}  -\{F,\Omega\}\{\Psi,G\}  \right) \,.
\end{eqnarray}
The associated  Hamiltonian $H_{\rm phys}$ is obtained from the total Hamiltonian after elimination of 
  $P$ and $\pi_\phi$ via the second-class constraints: 
\begin{eqnarray}
H_{\rm phys} = \frac{1}{2} (k^{-1})^{ij} p_ip_j + \frac{1}{2} \left( k_0 - (k^{-1})^{ij}c_ic_j \right) Q^2 + Q\,  v^iV_i  + V(\phi,q) \,.
\end{eqnarray}
The linear dependence on the canonical momentum $\pi_\phi$ has disappeared in the above Hamiltonian because of the constraint $\Psi$. This confirms that the  Ostrogradski ghost has been eliminated as a consequence of the degeneracy of the kinetic matrix $M$.

Note that our analysis focuses only on the Ostrogradski instability. Other types of instabilities could be present in the theory (such as, for instance, a ghost instability due to a negative eigenvalue of $k_{ij}$) and a further analysis of the above Hamiltonian, depending on the specific choice of the Lagrangian coefficients and of the potential,   is required to ensure the absence of any other dangerous instability.

\subsection{Summary}
The analysis of the Lagrangian (\ref{toy model action}) has  shown us that the crucial ingredient to avoid the presence of an Ostrogradski ghost is the degeneracy of the kinetic matrix (\ref{kinetic_matrix}). From the Hamiltonian point of view, this degeneracy entails the presence of  constraints,  which  reduce the number of physical degrees of freedom. The linear dependence of the Hamiltonian on one of the momenta, which is a signature of the Ostrogradski instability, is also eliminated.

In the rest of this paper, we will investigate the degeneracy of scalar tensor theories, following the same procedure as the one used in this section. The scalar field, with higher order derivatives, will be analogous to our variable $\phi(t)$, while the metric field  will play a r\^ole similar to that of the ``regular'' degrees of freedom $q^i(t)$. Of course, the mathematical structure is more complicated but the essential features concerning the degeneracy of the kinetic matrix turn out to be  quite similar.

In the toy model, a trivial way to ensure the degeneracy of the kinetic matrix is to impose $a=0$ and $b_i=0$.  As we will see, this situation is analogous to Horndeski theories where the equations of motion are second order\footnote{Horndeski Lagrangians contain terms linear in $\ddot\phi$, which we have not included in our toy model, but these terms are irrelevant for the kinetic matrix.}. 
Degeneracy can also be achieved in a non trivial way when  $b_i\neq 0$. The case $a=0$ and $b_i\neq 0$, which leads to third order equations of motion, is similar to the extension beyond Horndeski introduced in \cite{Gleyzes:2014dya}. We have also seen that the degeneracy condition can be satisfied even if $a\neq 0$. Remarkably, the same situation can occur in scalar-tensor theories, as we will see in Section~\ref{Section_classification}.

\section{Scalar tensor theories}
\subsection{The action}
We now consider a  class of scalar-tensor theories  whose dynamics is governed by an action of the general form
\beq\label{generalHorn}
S[g,\phi] \equiv \int \sqrt{\vert g \vert} \left( \f \, \R + C^{\mu\nu,\rho\sigma}\,  \nabla_\mu \nabla_\nu\phi \, \nabla_\rho \nabla_\sigma\phi \right)\,,
\eeq
where $\R$ is the four-dimensional Ricci scalar,  $\f$  an arbitrary function of $\phi$ and $X$ (a constant $\f$ corresponds to general relativity), and the tensor 
 $C^{\mu\nu,\rho\sigma}$ depends only on $\phi$ and $\phi_\mu\equiv \nabla_\mu\phi$.  
 One can also add in the above action other terms that depend only on $\phi_\mu$ or depend linearly on $\phi_{\mu\nu}$. Since these additional terms do not  modify the kinetic matrix, we will not need to consider them explicitly in the following in order to study the degeneracy of the Lagrangian.

Given the way it is contracted in the action,  one can require, without loss of generality, that $C^{\mu\nu,\rho\sigma}$
 satisfies the following symmetries: 
\beq
C^{\mu\nu,\rho\sigma} = C^{\nu\mu,\rho\sigma}= C^{\mu\nu,\sigma\rho}= C^{\rho\sigma,\mu\nu}\,.
\eeq 
As a consequence, it can always be written in the form  
\begin{eqnarray}\label{family}
C^{\mu\nu,\rho\sigma} & = &  \frac{1}{2} \ab\, (g^{\mu\rho} g^{\nu\sigma} + g^{\mu\sigma} g^{\nu\rho})+
\aa \,g^{\mu\nu} g^{\rho\sigma} +\frac{1}{2} \ac\, (\phi^\mu\phi^\nu g^{\rho\sigma} +\phi^\rho\phi^\sigma g^{\mu\nu} ) 
\cr
& & \quad +   \frac{1}{4} \ad (\phi^\mu \phi^\rho g^{\nu\sigma} + \phi^\nu \phi^\rho g^{\mu\sigma} + \phi^\mu \phi^\sigma g^{\nu\rho} + \phi^\nu \phi^\sigma g^{\mu\rho} ) +  \ae\, \phi^\mu \phi^\nu \phi^\rho \phi^\sigma\,, 
\label{four} 
\end{eqnarray}
where the $\a_i$ are functions of $\phi$ and $X$.

\subsection{Particular cases}
The class of theories (\ref{generalHorn}) includes as a particular case the quartic Horndeski term
\beq
L^{\rm H}_4 = \Gfour(\phi,X) \, \R - 2 \Gfour_{,X}(\phi,X) (\Box \phi^2 - \phi^{ \mu \nu} \phi_{ \mu \nu}) \,.
\eeq
The above Lagrangian is indeed of the form (\ref{generalHorn})-(\ref{family}) with 
\beq
f=\Gfour\,, \qquad \ab= -\aa= 2 \Gfour_{,X}\,, \qquad \ac=\ad=\ae=0\,.
\eeq

The action (\ref{generalHorn}) also includes the extension beyond Horndeski introduced in \cite{Gleyzes:2014dya}, which can be written as
\beq
L^{bH}_4=\Ffour(\phi,X) \epsilon^{\mu\nu\rho}_{\ \ \ \ \sigma}\, \epsilon^{\mu'\nu'\rho'\sigma}\phi_{\mu}\phi_{\mu'}\phi_{\nu\nu'}\phi_{\rho\rho'}\,.
\eeq
This corresponds to (\ref{generalHorn}) with 
\beq
\ab=-\aa= X \Ffour     \,, \qquad \ac=-\ad= 2 \Ffour\,, \qquad \ae=0\,.
\eeq

\subsection{Reformulation of the action}
Instead of working directly with  second order derivatives  in the Lagrangian,  it is more convenient to introduce a new variable, as illustrated in the toy model. In the present case, we simply replace in the action all first order derivatives $\nabla_\mu\phi$ by the components of 
a field $A_\mu$ and  we impose the relation $A_\mu = \nabla_\mu\phi$ via a constraint in the Lagrangian. Our new action is thus given by
\begin{eqnarray}\label{newform}
S[g,\phi;A_\mu,\lambda^\mu] = \int \sqrt{\vert g \vert} \left\{\f\,  {\R} + C^{\mu\nu,\rho\sigma} \nabla_\mu A_\nu \, \nabla_\rho A_\sigma + \lambda^\mu (\nabla_\mu\phi - A_\mu)\right\}\,,
\end{eqnarray}
where the tensor $C^{\mu\nu\rho\sigma}$ is now expressed in terms of $A_\mu$ and $\phi$.

It is not difficult  to verify  that (\ref{newform}) and (\ref{generalHorn}) are  equivalent at the classical level. 
To do so, let us write the equations of motion induced by (\ref{newform}) for the scalar field $\phi$ and the vector field $A_\mu$, 
\begin{eqnarray}
\frac{\delta C^{\mu\nu,\rho\sigma}}{\delta \phi} \nabla_\mu A_\nu \, \nabla_\rho A_\rho - \nabla_\mu \lambda^\mu = 0 \;\;\; ,�\;\;\;
\frac{\delta C^{\mu\nu,\rho\sigma}}{\delta A_\a} \nabla_\mu A_\nu \, \nabla_\rho A_\sigma -
 2\nabla_{\beta} \left( C^{\mu\nu,\a\beta} \nabla_\mu A_\nu\right) = \lambda^\a\,,
 \label{eq_A}
\end{eqnarray}
together with 
\beq
\label{A=Dphi}
A_\mu=\nabla_\mu\phi  \,,
\eeq
which follows from the variation with respect to  $\lambda^\mu$.
Taking the divergence of the last equation in (\ref{eq_A}) and replacing $A_\mu$ by $\nabla_\mu\phi$   leads to the usual equation of motion for the scalar field when
one uses the first equation. It is also immediate to check that the equations of motion for the metric are also equivalent. In the next section, we will use this new formulation  (\ref{newform})  to study the possible degeneracies of the Lagrangian.

\section{Degeneracy}
In this section, we concentrate on the kinetic part of the Lagrangian in order to write explicitly the degeneracy conditions for the kinetic matrix. We thus need to separate the time derivatives from the spatial derivatives. The standard procedure consists in a 3+1 decomposition {\it \`a la} ADM within  an explicit coordinate system. Here instead, we resort to a {\it covariant} 3+1 decomposition of spacetime, i.e. we do not introduce a coordinate system but work with tensors that are decomposed into  time-like and space-like components. For explicit calculations, this is much more efficient that the traditional ADM decomposition.
We use the abstract index notation, with latin indices ($a$, $b$, etc),  to emphasize that we are working directly with tensors.

\subsection{Kinetic matrix}
We  assume the existence of a slicing of spacetime with 3-dimensional spacelike hypersurfaces. We introduce their normal unit vector $n^a$, which is time-like, and satisfies the normalization condition $n_a n^a=-1$.  This induces a three-dimensional metric, corresponding to the projection tensor on the spatial hypersurfaces,  defined by
\beq
h_{ab}\equiv g_{ab}+n_a n_b\,.
\eeq
It is then useful to define the spatial projection of $A_a$,
 \beq
 \tA_a\equiv h_a^bA_b\,,
 \eeq
 and its normal projection
 \beq
 \An\equiv A_a n^a\,.
 \eeq

Let us now introduce the time direction vector $t^a=\partial/\partial t$ associated with a time coordinate $t$ that labels the slicing of spacelike hypersurfaces.
One can always decompose $t^a$ as 
\beq
t^a =N n^a +N^a,
\eeq
thus defining the lapse function $N$   and the shift vector $N^a$ orthogonal to $n^a$. We also define 
the ``time derivative'' of any  spatial tensor as the spatial projection of its Lie derivative  with respect to $t^a$. In particular, we have 
 \beq
 \dotAn \equiv t^a \nabla_a\An \,, \qquad \dot\tA_a\equiv h_a^b{\cal L}_t \tA_b=h_a^b\, (t^c \nabla_c \tA_b+\tA_c\nabla_b t^c)\,.
 \eeq

Using the above definitions, as well as the property $\nabla_a A_b=\nabla_b A_a$ which follows from (\ref{A=Dphi}), one finds that  the 3+1 covariant decomposition of  $\nabla_a A_b$ is given by\footnote{In particular, we have used the relation $\dot\tA_b=D_b(A_c t^c)+\An(N a_b-D_bN)$, which is valid when $\nabla_{[a}A_{b]}=0$, to eliminate the time derivatives of $\tA_b$.}
 \begin{eqnarray}
 \label{decomposition_DA}
 \nabla_a A_b&=&D_a\tA_b-\An K_{ab}
 +n_a ( K_{bc}\tA^c    -D_b\An)
 +n_b (K_{ac}\tA^c - D_a\An) 
 \cr
 &&
 +\frac1N n_a n_b \, (\dotAn - N^c D_c\An -N \tA_c a^c)\,,
 \end{eqnarray}
 where $D_a$ denotes the 3-dimensional  covariant derivative associated with the spatial metric $h_{ab}$, $a_b\equiv n^c\nabla_c n_b$ is the ``acceleration'', 
 and 
 $K_{ab}$ is the extrinsic curvature tensor, which 
can be expressed as 
\beq
\label{Kab}
K_{ab}=\frac{1}{2N}\left(\dot h_{ab}-D_a N_b -D_b N_a\right)\,.
 \eeq

The only terms in (\ref{decomposition_DA}) that  are  relevant for  the kinetic part of the Lagrangian are
\beq
(\nabla_a A_b)_{\rm kin}=\lambda_{ab}\, \dotAn+\Lambda_{ab}^{\ \ cd} \, K_{cd}\,,
\eeq
where we have introduced the tensors
\beq
\label{lambda}
 \lambda_{ab}\equiv \frac1N n_a n_b\,, \qquad \Lambda_{ab}^{\ \ cd}=-\An \, h_{(a}^c h_{b)}^d+2\, n_{(a} h_{b)}^{(c} \tA^{d)}\,.
 \eeq
 Strictly speaking, only the $\dot h_{ab}$ term is relevant but we will keep $K_{ab}$ for convenience. 
 We thus find that the kinetic part of the Lagrangian quadratic in $\nabla_a A_b$ reduces to 
 \beq
 L^{(\phi)}_{\rm kin}=C^{ab,cd}\lambda_{ab}\, \lambda_{cd} \, \dotAn^2+2 C^{ab,cd}\Lambda_{ab}^{\ \ ef}\lambda_{cd}\, \dotAn  K_{ef}
  + C^{ab,cd}\Lambda_{ab}^{\ \ ef}\Lambda_{cd}^{\ \ gh} K_{ef}K_{gh}\,,
 \eeq
 which is similar to the Lagrangian (\ref{reformulated toy model}), with $\An$ and $K_{ab}$ (or $\dot h_{ab}$)  playing  the r\^ole of $Q$ and $\dot q_i$, respectively.

 One can then compute the analogs of the coefficients $a$, $b_i$ and  $k_{ij}$ in (\ref{reformulated toy model}),  up to a factor $2$ for notational convenience, 
   by substituting the explicit expressions for $C^{ab,cd}$, $\lambda_{ab}$ and $\Lambda_{ab}^{\ \ cd}$. After straightforward calculations, we  find that 
 the first kinetic coefficient is given by
 \beq
 \label{A}
\A\equiv C^{ab,cd}\lambda_{ab}\lambda_{cd} =\frac{1}{N^2}\left[\ab+\aa-(\ac+\ad)\An^2+
 \ae \An^4\right]
 \,,
 \eeq
while the coefficients of the mixed terms can be written as 
 \beq
 {\cal B}^{ef}\equiv C^{ab,cd}\Lambda_{ab}^{\ \ ef}\lambda_{cd}
 =\ba h^{ef} +\bb \,  \tA^e\tA^f\,,
 \label{B_decomp}
 \eeq
 with
 \beq
 \ba=
 \frac{\An }{2N}\left(2\aa-\ac\An^2\right)  \,, \qquad 
 \bb=-\frac{\An }{2N}\left(\ac+2\ad-2\ae\An^2\right)  \,.
 \eeq
 
 Finally the kinetic coefficient for the purely metric  sector is given by 
 \beq
 {\cal K}^{ef,gh}\equiv C^{ab,cd}\Lambda_{ab}^{\ \ ef}\Lambda_{cd}^{\ \ gh}\,.
 \eeq
 Substituting the explicit expressions in (\ref{family}) and (\ref{lambda}), one gets
 \begin{eqnarray}
  {\cal K}^{ab,cd}\equiv  C^{ab,cd}\Lambda_{ab}^{\ \ ef}\Lambda_{cd}^{\ \ gh} &=&\kb h^{a(c}h^{d)b}
 +\ka\, h^{ab} h^{cd}+
  \frac12 \kc\left(\tA^a\tA^b h^{cd}+\tA^c \tA^d h^{ab}\right)
    \cr
  &&
  +\frac 12 \kd  \left(\tA^a\tA^{(c}h^{d) b}+\tA^b\tA^{(c} h^{d)a}\right) 
  +\ke \tA^a \tA^b\tA^c\tA^d\,,
  \label{K_decomp}
 \end{eqnarray}
 with 
 \begin{eqnarray}
 \kb=\ab \An^2\,, \quad\ka=  \aa \An^2\,, \quad    \kc=- \ac\An^2\,, \quad  \kd=- 2\ab\,,\quad  \ke=\ae\An^2-\ad\,.
 \end{eqnarray}
 One can note that the structure of ${\cal K}^{ab,cd}$ is completely analogous to that  of $C^{ab,cd}$, the only difference being that the former depends on the spatial metric $h_{ab}$ and spatial vector $\tA_a$, whereas the latter depends on the spacetime metric $g_{ab}$ and vector $A_a$.

 To obtain the full kinematic part of the action, one must also take into account  the gravitational term $f \R$. Using the identity  
 \beq
 \R=K_{\mu\nu}K^{\mu\nu} -K^2 +\Rs-2\nabla_\mu (a^\mu-K n^\mu)\,,
 \eeq
  and   integrating by parts, one can rewrite the gravitational part of the action as 
 \beq
 \label{Lag_grav}
 \int d^4x \sqrt{-g} \, \f \, \R=\int d^4x \sqrt{-g}\left\{f \left[K_{\mu\nu}K^{\mu\nu} -K^2 +\Rs\right]
 +2 \nabla_\mu f \left(a^\mu -K n^\mu\right)\right\}\,,
 \eeq
 where $\Rs$ is the three-dimensional Ricci scalar. 
 Since $\nabla_\mu f= 2f_X A^\nu A_{\mu\nu}+f_\phi A_\mu$, one finds that the second term on the right hand side contributes to the mixed kinetic terms $\dot\An K_{ab}$, with the coefficient 
 \beq
 \label{Bgrav}
 {\cal B}^{ab}_{\rm grav}=2\f_X\frac{\An}{N} \, h^{ab}\,.
 \eeq
 The two terms in (\ref{Lag_grav}) contribute to the kinetic term quadratic in $K_{ab}$ and the corresponding coefficient can be written as
 \beq
 {\cal K}^{ab,cd}_{\rm grav}=\gb h^{a(c}h^{d)b} 
 +  \ga\,  h^{ab}h^{cd}+\frac12 \gc \left(\tA^a\tA^b h^{cd}+\tA^c \tA^d h^{ab}\right)\,,
 \eeq
with 
\beq
\gb=-\ga=\f\,, \qquad \gc=4 \f_X \,.
\eeq
In summary the coefficients obtained from the total action are 
\beq
\tilde {\cal B}^{ab}= {\cal B}^{ab}+ {\cal B}^{ab}_{\rm grav},\, \qquad
\tilde {\cal K}^{ab,cd}={\cal K}^{ab,cd}+{\cal K}^{ab,cd}_{\rm grav}\,,
\eeq
which can be decomposed as in (\ref{B_decomp}) and (\ref{K_decomp}), respectively, with the new coefficients 
$\tba$, $\tbb$, $\tka$, $\tkb$ and $\tkc$, while the coefficients $\kd$ and $\ke$ remain unchanged.  The coefficients $\A$, $\tilde {\cal B}^{ab}$ and $\tilde {\cal K}^{ab,cd}$ play the same r\^ole as, respectively,  $a$, $b_i$ and $k_{ij}$ in the toy model.

Interestingly, in the case of the Horndeski Lagrangian $L_4^{\rm H}$, we have 
\beq
\ba=-2\frac{A_*}{N} G_{4X}, \qquad \bb=0\,,\qquad ({\rm Horndeski})
\eeq
and one notes that ${\cal B}^{ab}$ is exactly cancelled by the gravitational contribution (\ref{Bgrav}), so that the total coefficient $\tilde {\cal B}^{ab}$ vanishes. This is not surprising since Horndeski's theories are, by construction,  restricted to give second order equations of motion. By contrast, when $\tilde {\cal B}^{ab}\neq 0$, the equations of motion become higher order, as with $L_4^{\rm bH}$. 

 \subsection{Degeneracy conditions}
 \subsubsection{No dynamical metric}
 Let us start with the extremely simple situation where only the dynamics of the scalar field is taken into account, while the spacetime metric is frozen. In this case, the kinetic Lagrangian is reduced to the $\dotAn^2$ term and the system is degenerate only if $\A=0$. 
Since it must be true independently of the particular value of $\An$, this implies, according to (\ref{A}), the three conditions
\beq
\label{no-ghost}
\ab+\aa=0\,,  \qquad \ac+\ad=0\,, \qquad  \ae=0\,.
\eeq
We note that both quartic Lagrangians $L_4^{\rm H}$ and $L_4^{\rm bH}$ satisfy these conditions. This is expected since both Lagrangians reduce  in flat spacetime to galileons, which are not plagued with ghosts. 
 
 \subsubsection{Dynamical metric}
 In the general case, one must now consider the full kinetic matrix, which is of the form
\beq
\left(
\begin{array}{cc}
\A &\tilde\B^{cd}\\
\tilde \B^{ab} & \tilde {\cal K}^{ab,cd}
\end{array}
\right)\,.
\eeq
This matrix is degenerate if there exists an eigenvector with zero eigenvalue, i.e. if  one can find $v_0$ and $\V_{cd}$ such that 
\beq
\label{degeneracy}
v_0 \, \A+\tilde\B^{cd} \V_{cd}=0\,,  \,\qquad  v_0 \, \tilde\B^{ab}+ \tilde{\cal K}^{ab,cd} \, \V_{cd}=0\,.
\eeq
Since $\V_{cd}$ is a symmetric spatial tensor of order 2, it must be of the form 
\beq
\V_{cd}=v_1 \, h_{cd}+ v_2\,  \tA_c\tA_d\,,
\eeq
and the contraction of ${\cal K}^{ab,cd}$ with $\V_{cd}$ can be similarly decomposed along $h^{ab}$ and $\tA^a\tA^b$. In this way, the system (\ref{degeneracy}) is easily transposed into the matricial relation
%%% corrected
\beq
\M\cdot \V\equiv
\left(
\begin{array}{ccc}
{\cal A} \quad & 3\tilde\beta_1+\tilde\beta_2 \hat A^2  \quad & \tilde\beta_1 \hat A^2+\tilde\beta_2 (\hat A^2)^2
\\
\tilde\beta_1 \quad &  \tilde\kappa_1+3\tilde\kappa_2+\tilde\kappa_3 \hat A^2/2 \quad &  \tilde\kappa_2 \hat A^2+\tilde\kappa_3 (\hat A^2)^2/2
\\
\tilde\beta_2 \quad  & 3\tilde\kappa_3/2+\kappa_4 +\kappa_5 \hat A^2  \quad  & \tilde\kappa_1+ (\tilde\kappa_3/2+\kappa_4)\hat A^2+\kappa_5 (\hat A^2)^2 
\end{array}
\right)
\left(
\begin{array}{c}
v_0 \\ v_1 \\ v_2
\end{array}
\right)=0\,.
\eeq
%%%
There is a nontrivial solution if the determinant of the matrix $\M$ vanishes. The matrix components depend on the functions $\f$ and $\a$'s, as well as on $\An$ and $\tA^2$, and  the latter quantity can be  expressed in terms of $\An$ and $X$, since $\tA^2=X+\An^2$. 

Requiring the determinant of $\M$ to vanish yields an expression of the form 
\beq
\label{determinant}
\D_0(X)+\D_1(X) \An^2+\D_2(X) \An^4=0\,,
\eeq
with
\beq
\D_0(X)\equiv -4 (\aa+\ab) \left[X \f (2\ab+X\ad+4\f_X)-2\f^2-8X^2\f_X^2\right]\,,
\eeq
\begin{eqnarray}
\D_1(X)&\equiv& 4\left[X^2\ab (\ab+3\aa)-2\f^2-4X\f \aa\right]\ad +4 X^2\f(\ab+\aa)\ae +8X\ab^3
\cr
&&
-4(\f+4X\f_X-6X\aa)\ab^2 -16(\f+5X \f_X)\ab \aa+4X(3\f-4X \f_X) \ab\ac 
\cr
&&
-X^2\f \ac^2 +32 \f_X(f+2X f_X) \aa-16\f \f_X \ab-8\f (\f-X\f_X)\ac+48\f \f_X^2 \,,
\end{eqnarray}
\begin{eqnarray}
\D_2(X)&\equiv& 4\left[ 2\f^2+4X\f \aa-X^2\ab(\ab+3\aa)\right]\ae  + 4\ab^3+4(2\aa-X\ac-4\f_X)\ab^2+3X^2 \ab\ac^2
\cr
&&
-4X\f \ac^2+8 (\f+X\f_X)\ab\ac -32 \f_X \ab\aa+16\f_X^2\ab
+32\f_X^2\aa-16\f\f_X\ac\,.
\end{eqnarray}
Since the determinant must vanish for any value of $\An$, we deduce that degenerate theories are characterized by the three conditions
\beq
\D_0(X)=0, \qquad \D_1(X)=0, \qquad \D_2(X)=0\,.
\eeq
We identify in the next section the theories of the form (\ref{generalHorn}) that satisfy  these conditions simultaneously.

\section{Classification of degenerate theories}
\label{Section_classification}
The  condition $D_0(X)=0$ is the simplest of all three and allows to distinguish two subclasses of theories, depending on whether the condition $\ab+\aa=0$ is satisfied or not. Note that 
this condition is also one of the conditions to get $\A=0$.

\subsection{Models with $\ab+\aa=0$}
In this case $D_0(X)=0$ is automatically satisfied.  
One can then use the condition  $\D_1(X)=0$ to express $\ad$ in terms of $\aa$ and $\ac$:
\begin{eqnarray}
\ad&=&\frac{1}{8(f+X\aa)^2}\left[16 X \aa^3+4 (3\f+16 X\f_X)\aa^2
+(16X^2 \f_X-12X\f) \ac\aa-X^2\f \ac^2
\right.
\cr
&&\qquad\qquad \qquad 
\left.
+16 \f_X(3\f+4X\f_X)\aa+8\f (X\f_X-\f)\ac+48\f \f_X^2\right]\,.
\end{eqnarray}
Similarly, the condition $D_2(X)=0$ yields
\beq
\ae=\frac{\left(4\f_X+2\aa+X\ac\right)\left(-2\aa^2+3X\aa\ac-4\f_X \aa+4\f \ac\right)}{8(f+X\aa)^2}\,.
\eeq
We thus conclude that degenerate theories in this subclass depend on three arbitrary functions $\aa$, $\ac$ and $f$. 

Focussing now on theories satisfying $\A=0$, which imposes the additional conditions $\ae=0$ and  $\ac+\ad=0$, one finds  that the functions $\aa$ and $\ac$ are no longer independent, but related by
\beq
\label{cond2}
4\f_X+2\aa+X\ac=0\,.
\eeq
This means that the condition $\A=0$ restricts the degenerate theories to a subclass that depends on two arbitrary functions only. 

It is easy to see that this family of theories in fact coincides with the sum of $L_4^{\rm H}$ and  $L_4^{\rm bH}$, upon using  the identification
\beq
\f=G_4\,,\qquad  \ab=-\aa=2 G_{4X}+ X F_4\,, \qquad \ac=-\ad=2 F_4\,.
\eeq
This implies that the quartic Lagrangian $L_4=L_4^{\rm H}+L_4^{\rm bH}$ represents  the most general theories of the form (\ref{generalHorn}) that are  degenerate both with or without gravity.

Note that the contribution from $L_4^{\rm bH}$ to the coefficient $\B^{ab}$ is given by
\beq
\B^{ab}_{_{4{\rm bH}}}=\frac{\An}{N}F_4 (\tA^a\tA^b-\tA^2 h^{ab})\,,
\eeq
whereas
\beq
{\cal K}^{ab,cd}_{_{4{\rm bH}}}\tA_c\tA_d=F_4 \An^2\,  (2\tA^2-\An^2) (\tA^a\tA^b-\tA^2 h^{ab})\,,
\eeq
which means that the kinetic matrix of $L_4^{\rm bH}$ has a null eigenvector characterized by $\V_{cd}=\tA_c\tA_d$ and $v_0=-N\An (2\tA^2-\An^2) $. When we consider the sum $L_4=L_4^{\rm H}+L_4^{\rm bH}$, one finds 
\beq
\tilde\B^{ab}_4=\B^{ab}_{_{4{\rm bH}}}\,,
\eeq
and 
\beq
\tilde{\cal K}^{ab,cd}_4\tA_c\tA_d=\left[G_4-2XG_{4X}+F_4 (2X+\An^2)\An^2\right] (\tA^a\tA^b-\tA^2 h^{ab})\,,
\eeq
which shows that $L_4^{\rm H}+L_4^{\rm bH}$ is degenerate with the null eigenvector defined by $\V_{cd}=\tA_c\tA_d$ and $v_0=-N [G_4-2XG_{4X}+F_4 (2X+\An^2)\An^2]/(\An F_4)$.

\subsection{Models with $\ab+\aa\neq 0$}
To simplify the presentation, we now assume that the gravitational sector is described by general relativity, i.e. $f=1$, but it  is immediate to extend the results given below to the general case.
In this situation, the first condition, $\D_0(X)=0$,  is satisfied if 
\beq
\label{cond3}
2X\ab+X^2\ad=2 \,.
\eeq
We can then proceed as in the previous subsection by solving $D_1(X)=0$ and $D_2(X)=0$ to express $\ad$ and  $\ae$ in terms of the three other functions. Substituting the obtained expression for $\ad$ into the condition
(\ref{cond3}),  one finally gets
\beq
(X\ab -1)^2 (4 + 8X \aa  + 2X \ab  + X^2\ac)^2=0\,.
\eeq
We thus have two subcases. In the first subcase ($\ab =1/X$), we find the family
\beq
 \ab=\frac1X\,,\qquad \ad=0\,,\qquad \ae=\frac{-4 - 8X \aa  - 4 X^2\ac  + X^4\ac^2 }{4 X^3 (1 + X\aa )}\,,
\eeq
where $\aa$ and $\ac$ are arbitrary functions.
In the second subcase, we get
\beq
 \ab=-\frac2X-4 \aa -\frac{X}2 \ac\,,\qquad \ad=\frac6{X^2}+\frac8X\aa+\ac,\qquad \ae=-\frac{4 + 8 X\aa  +3X^2 \ac }{X^3}\,, 
\eeq
while  $\aa$ and $\ac$ are arbitrary.

It is straightforward to repeat the same calculation with an arbitrary $\f$. We have not written down the results we got because the expressions are a bit cumbersome  and not very illuminating. 
In conclusion of this subsection, we have obtained two other families of theories, which  depend on the three arbitrary functions $\aa$, $\ac$ and $\f$.

\section{Quintic Lagrangian beyond Horndeski}
In the previous sections, we have systematically investigated the Lagrangians of the form (\ref{generalHorn}), which include the quartic terms $L_4^{\rm H}$ and $L_4^{\rm bH}$. A similar systematic investigation of the Lagrangians with a cubic dependence on the second derivatives of $\phi$ is much more involved and is left for future work. In this section, we  just consider the quintic Lagrangian beyond Horndeski, i.e.  $L_5^{\rm bH}$, whose dependence on $\phi_{ab}$ is cubic:
\beq
L_5^{\rm bH} = C^{ab,cd,ef}_{_{5\rm bH}} \phi_{ab} \,\phi_{cd} \, \phi_{ef}\,.
\eeq
The tensor $C^{ab,cd,ef}$ satisfies the following symmetries: invariance under the exchange of $a$ and $b$, of $c$ and $d$, and of $e$ and $f$, and invariance under permutations of the pairs $(ab)$, $(cd)$ and $(ef)$. According to (\ref{L5bH}), this tensor is given explicitly by
\beq
C^{ab,cd,ef}_{_{5\rm bH}}=F_5\, {\rm Sym}_{\substack{\{\{a,b\},\{c,d\},\{e,f\}\}}} \left[\phi_{a'}\,\phi_{b'}\, \epsilon^{a'a\,c\,e}\,\epsilon^{b'b\,d\,f}\right]\,,
\eeq
where we symmetrize the expression between the brackets with respect to the index symmetries listed above.

We now derive  the matrix consisting of the second derivatives of the Lagrangian with respect to the velocities $\dotAn$ and $K_{ab}$.  By analogy with the previous sections, this matrix  will also  be called the kinetic matrix, although it now depends on the velocities. 
The contribution to ${\cal A}$ from $L_5^{\rm bH} $ is given by
% corrected
\beq
\A_{_{5\rm bH}}\equiv\frac12\, \frac{\partial^2 L_5^{\rm bH}}{\partial\dotAn^2}=\frac12\, \frac{\partial^2 L_5^{\rm bH}}{\partial \phi_{ab}\,\partial \phi_{cd}}\lambda_{ab} \lambda_{cd}=3\, C^{ab,cd,ef}_{_{5\rm bH}} \,\lambda_{ab} \,\lambda_{cd}\, \phi_{ef}\,,
\eeq
which always vanishes because 
\beq
C^{ab,cd,ef}_{_{5\rm bH}} \,\lambda_{ab} \,\lambda_{cd}=0\,.
\eeq
Similarly, the contribution of  $L_5^{\rm bH} $ to $\B^{ab}$ is
\begin{eqnarray}
\B^{ab}_{_{5\rm bH}}&\equiv&\frac12\, \frac{\partial^2 L_5^{\rm bH}}{\partial\dotAn\partial K_{ab}}=3\, C^{cd,ef,gh}_{_{5\rm bH}} \,\lambda_{cd} \,\Lambda_{ef}^{\ \ \ ab} \,\phi_{gh}
\\
&=& 3 F_5 \frac{\An}{N}\left[\tA^2 \Sigma^{ab}+\left(\tA^e\tA^f\Sigma_{ef}-\tA^2\Sigma\right)h^{ab}-2\tA_e\Sigma^{e(a}\tA^{b)}
+\Sigma\tA^a\tA^b\right]
\,,
\end{eqnarray}
where $\Sigma_{ab}\equiv D_a \tA_b-\An K_{ab}$ is the purely spatial part of $\phi_{ab}$ (see Eq.~(\ref{decomposition_DA})), and $\Sigma=h^{ab}\Sigma_{ab}$ its trace. It is immediate to check the identity
\beq
\label{B5AA}
\B^{ab}_{_{5\rm bH}}\tA_a \tA_b=0\,,
\eeq
which will be useful below to identify a null eigenvector of the 
kinetic  matrix.

Finally,  the last contribution to the 
kinetic  matrix reads
\beq
{\cal K}^{ab,cd}_{_{5\rm bH}}\equiv\frac12\, \frac{\partial^2 L_5^{\rm bH}}{\partial K_{ab}\partial K_{cd}}=3\, C^{ef,gh,mn}_{_{5\rm bH}} \,\Lambda_{ef}^{\ \ \ ab} \,\Lambda_{gh}^{\ \ \ cd} \,\phi_{mn}\,,
\eeq
which implies, in particular, 
\beq
{\cal K}^{ab,cd}_{_{5\rm bH}}\tA_c\tA_d= N \An (2\tA^2-\An^2) \B^{ab}_{_{5\rm bH}}\,.
\eeq
This relation, together with (\ref{B5AA}), shows that the kinetic matrix of $L_5^{\rm bH}$ obeys  the degeneracy conditions (\ref{degeneracy}) with $\V_{ab}=\tA_a \tA_b$ and $v_0= -N \An (2X+\An^2)$.
This proves that the Lagrangian beyond Horndeski $L_5^{\rm bH}$ is degenerate. Moreover, the eigenvector is the same as for $L_4^{\rm bH}$, which implies that the combination of $L_4^{\rm bH}$ and $L_5^{\rm bH}$ is also degenerate. 

However, the combination of $L_4^{\rm H}$, $L_4^{\rm bH}$ and $L_5^{\rm bH}$  does not yield a degenerate Lagrangian, because the null eigenvectors do not coincide, except if $G_4-2X G_{4X}=0$. Note that, in this respect, the unitary gauge (i.e. with a uniform  scalar field  on constant time hypersurfaces)  is somewhat misleading since the coefficients $\B^{ab}$ for the Lagrangians $L_4^{\rm H}$, $L_4^{\rm bH}$ and $L_5^{\rm bH}$ all vanish in this gauge ( in which $\tA^a=0$), which suggests that the sum of all these Lagrangians is degenerate.  As the present analysis shows,  this is in fact not the case.  Other examples where the unitary gauge can be misleading are discussed in the Appendix.

\section{Conclusions}
In this work, we have studied a  large class of higher derivative scalar theories coupled to gravity, which include the quartic Horndeski  Lagrangian $L^{\rm H}_4$ and its extension beyond Horndeski $L_4^{\rm bH}$.
 As we have shown, the coupling of the scalar field to other degrees of freedom  extends the range of possibilities to construct degenerate theories, in order to avoid Ostrogradski's ghosts. We have illustrated these new possibilities with a simple toy model, which is easy to analyse  in both the Lagrangian and Hamiltonian formulations. 

In summary, one can distinguish three possibilities. The simplest one corresponds to a trivial degeneracy where the kinetic matrix has a row (and column) of zeros, thus yielding second order  equations of motion. Horndeski's theories, which were required by construction to give second order equations of motion, are examples of this simple case. 
The second possibility  occurs when the degeneracy of the kinetic matrix is nontrivial as a consequence of the coupling with the other degrees of freedom, but  remains degenerate when  this coupling is suppressed. This is what happens with  the extension beyond Horndeski $L_4^{\rm bH}$, leading to third order equations of motion. 
Finally, we also have  the possibility that the degeneracy is entirely due to the couplings with the other degrees of freedom and disappears when the latter are suppressed. 

For our toy model, we have fully derived the Hamiltonian formulation and shown explicitly the link between the degeneracy of the kinetic matrix and the absence of the Ostrogradski instability. From the Hamiltonian point of view, the disappearance of the ghost-like degree of freedom is the direct consequence of the presence of two second-class constraints in phase space\footnote{Note that the degeneracy obtained via the coupling to other degrees of freedom naturally provides these constraints in our models, in contrast with  the models discussed in \cite{Chen:2012au} where the constraints are imposed by hand.}. 

The derivation of a Hamiltonian formulation for scalar tensor theories of the form (\ref{generalHorn}) in an arbitrary gauge will be presented in another publication~\cite{Langlois:2015skt}. The Hamiltonian analysis in an arbitrary gauge is much more involved than in the unitary gauge, which was used in  \cite{Gleyzes:2014dya,Lin:2014jga,Gleyzes:2014qga}, and so far, only the Lagrangian $L_4^{\rm bH}$ has been considered, with the conclusion  that the number of degrees of freedom is strictly less than four in this particular case~\cite{Deffayet:2015qwa}.
Our  Hamiltonian formulation, detailed   in \cite{Langlois:2015skt},  is based on the tools developed in the present work  and confirms the absence of an extra degree of freedom in  the degenerate theories that we have identified.

The analysis of the present work is mainly devoted to theories that are quadratic in the second derivatives of $\phi$. A systematic treatment of the theories that are cubic in $\phi_{\mu\nu}$, which include  $L_5^{\rm bH}$, is more involved, because the coefficients of the kinetic matrix keep a linear dependence on $\phi_{\mu\nu}$,  and is left for future work. 
We have nevertheless checked explicitly that the kinetic matrix of the Lagrangian $L_5^{\rm bH}$ is also degenerate. Its direction of degeneracy is the same  as that of $L_4^{\rm bH}$, but not that of the sum $L_4^{\rm H}+L_4^{\rm bH}$ in general. This means that only restricted combinations of Horndeski's Lagrangians with the extensions $L_4^{\rm bH}$ or $L_5^{\rm bH}$   lead to degenerate theories and are thus presumably free of Ostrogradski instabilities.
Note that these findings are consistent with the results of \cite{Gleyzes:2014qga} showing the correspondence between theories beyond Horndeski and Horndeski theories via disformal transformations for restricted subclasses of Lagrangians. 
Moreover, these conclusions should not be seen as inconsistent with the property that  the equations of motion for a general combination of Horndeski and beyond Horndeski terms can be rewritten as a system of equations which are second-order in time derivatives, as shown in \cite{Deffayet:2015qwa}. Indeed, the redundancy of variables that describe the two gravitational degrees of freedom obscures the relation between the order of the equations of motion and the number of degrees of freedom.

Beyond shedding some light on the structure of Horndeski's theories and of their extensions, our systematic treatment of the degeneracy provides  a new tool for identifying ghost-free theories, much easier than a full Hamiltonian analysis. In the present context, this approach has enabled us  to identify other classes of scalar-tensor Lagrangians. In contrast with Horndeski and their extensions, these Lagrangians have the property to be  non degenerate when the metric is non dynamical, and thus suffer from Ostrogradski instabilities in this case. Remarkably however,  the Lagrangian becomes degenerate,  via the  coupling between the scalar field and the metric kinetic terms, when the metric is dynamical. This suggests that the Ostrogradski ghost is tamed by gravity for these theories. We leave for future work a more detailed investigation of these theories.

\acknowledgements
We would like to thank Christos Charmousis and Gilles Esposito-Farese for instructive discussions. 
DL thanks the hospitality of the Aspen Center for Physics, where some of the research related to this work was carried out. 
DL also thanks the organizers of the workshop ``Exploring Theories of Modified Gravity'', KICP, University of Chicago (October 2015) and the participants of this workshop for interesting discussions about this work.

\appendix

\section{Unitary gauge}
Instead of following the procedure outlined in the main text, one might be tempted to work directly in the so-called unitary gauge, where the constant time hypersurfaces coincide with uniform scalar field hypersurfaces. In our language, this gauge is characterized  by the condition
\beq
\tA_a=0\,.
\eeq
In order to determine the degeneracy of the kinetic matrix in the unitary gauge, one can  proceed as in the general case. In this way, one obtains a matrix $\M_u$ that simply corresponds to the matrix $\M$ of the main text restricted to the case $\tA=0$, i.e.
\beq
\M_u=\M_{{|\tA^2=0}}\,.
\eeq
Moreover, since $X=-\An^2$ in the unitary gauge, the condition det$(\M_u)=0$ yields an expression of $X$ only, given by
\beq
\label{det_unitary}
\D_0(X)-X\D_1(X) +X^2\D_2(X) =0\,,
\eeq 
where we have simply replaced $\An^2$ by $-X$ in the expression (\ref{determinant}).

This result shows that one must be careful when using the unitary gauge to determine whether a theory is degenerate or not.   Indeed, a theory can obey the above condition (\ref{det_unitary}) without satisfying the three conditions $D_i(X)=0$ separately. 
A simple example of this type,  which looks degenerate in the unitary gauge but turns out to be non-degenerate, is described by
\beq
\ab=-X\ad\,, \qquad \aa=\ac=\ae=0\,.
\eeq
For $\ad=-1/2$, the corresponding  Lagrangian simply reads\footnote{We thank Gilles Esposito-Farese for giving us, at an early stage of this work,  this simple example which illustrates some potential pitfall of the unitary gauge.}
\beq
L=\nabla_{[\lambda}\phi \ \nabla_{\mu]}\!\nabla_\nu \phi \  \nabla^{[\lambda}\phi \  \!\nabla^{\mu]}\nabla^\nu \phi\,.
\eeq

\end{document}